# Evanescent Field Functional $Cu_{3-x}P$ Nanoparticles as Effective Saturable Absorbers with high Repeatability for Femtosecond Soliton Pulse generation

*Haoran Mu[1], Zeke Liu[2], Zhichen Wan[1], Babar Shabbir[1], Lei Li[3], Tian Sun[2], Shaojuan Li[2], Wanli Ma[2], and Qiaoliang Bao[1,*].*

[1]Department of Materials Science and Engineering and ARC Centre of Excellence in Future Low-Energy Electronics Technologies (FLEET), Monash University, Clayton, Victoria 3800, Australia

[2]Institute of Functional Nano and Soft Materials (FUNSOM), Jiangsu Key Laboratory for Carbon-Based Functional Materials and Devices, and Collaborative Innovation Center of Suzhou Nano Science and Technology, Soochow University, Suzhou 215123, P. R. China.

[3]Jiangsu Key Laboratory of Advanced Laser Materials and Devices, Jiangsu Collaborative Innovation Center of Advanced Laser Technology and Emerging Industry, School of Physics and Electronic Engineering, Jiangsu Normal University, Xuzhou, Jiangsu 221116, China

*Address correspondence to (Q. Bao) Qiaoliang.Bao@ monash.edu

**Abstract**

Recently, a new emerging field about heavily-doped colloidal plasmonic nanocrystals (NCs) has attracted great attention due to their lower and expediently adjustable free carrier densities, lower and tunable LSPR band in the spectral range from NIR to MIR and higher optical nonlinearity. These new kinds of plasmonic materials will show huge potential and opportunities for nonlinear optical applications, such as ultrafast switching, nonlinear sensing and pulse laser generation. In this work, we demonstrate that high-quality mode-locking and Q-switching pulses at 1560 nm can both be generated by using controllable concentration of $Cu_{3-x}P$ NCs solution and fabricating evanescently interacted saturable absorbers. Furthermore, the plasmonic NCs material has good reproduction for fabricating SA devices and promising potential for large-scale industrial production. Our results may attract great attention for further investigations of heavily-doped plasmonic NCs as next generation, cheap and solution-processed element for fascinating applications in optoelectronic devices.

**Introduction**

Passively mode-locked fiber lasers, owing to their simple and compact structure, excellent light-beam quality, cost effectivity, and good compatibility to fiber optical systems, has achieved worldwide applications in different fields including fiber communications, materials machining, medical surgery, and so on.[1] In order to realize the passive mode-locking operation, the nonlinear optical materials called saturable absorbers (SA) would be inserted into the laser system without any additional mechanical or electrical modulation.[2] Due to the selective absorption between strong and weak light of the SA, the longitude mode of the laser would be locked and the output pulses would be compressed at the sub-picosecond level. Generally, bulk materials have weak optical nonlinearity which hardly meets the practical mode-locking application, therefore semiconductor quantum wells called semiconductor saturable absorber mirrors (SESAM) were introduced. [3] However, they have many drawbacks such as complex and expensive fabrication processes. In recent years, different low-dimensional nanomaterials which have inherent quantum confinement effects and large optical nonlinearity have achieved excellent mode-locking performances.[4-8] For example, the graphene based mode-locked fiber laser with the ultra-compact laser cavity (~ 1 cm) generated subpicosecond pulses[9], while both $WS_2$ and black phosphorus have achieved sub-150 fs pulse generation in fiber laser systems[10, 11]. However, the nanomaterials based saturable absorption devices are still far away from the commercial applications because most of these devices have low repeatability and are difficult to fabricate on large scales.

Heavily-doped colloidal plasmonic nanocrystals (NCs) such as self-doped copper-based chalcogenide NCs ($Cu_{2-x}S$, $Cu_{2-x}Se$, $Cu_{2-x}Te$), extrinsically-doped NCs, tungsten oxide and germanium telluride, have long been fascinated due to their unique localized surface plasmonic resonance (LSPR) properties[12-14]. Firstly, unlike the most intensively investigated noble plasmonic metals such as gold, silver and copper, in which plasmonic absorption focuses on visual waveband, the LSPR frequency of heavily-doped NCs is commonly located at infra-waveband due to their lower carrier concentrations.[15, 16] Secondly, through adjusting the composition of the elements, their LSPR frequency can be tuned drastically, which indicates the controllable working wavelength.[17] Owing to the nonlinear optical enhancement at LSPR frequency, the plasmonic semiconductor NCs have potential to be excellent SAs for mode-locking generation.[18] In 2015, self-doped colloidal copper phosphide ($Cu_{3-x}P$) NCs as a typical member of self-doped colloidal plasmonic NCs were firstly investigated to have ultrafast plasmonic dynamic response (~130 fs of the fast plasmonic relaxation component $\tau_1$) and huge optical nonlinearities of plasmonic resonances (> 18% of the light modulation depth).[14] By drop-casting its solution onto the fiber endfacet as a new kind of saturable absorber, high-energy and stable Q-switching pulses were also generated in a 1.5 µm fiber laser. Later, $Cu_{2-x}S$ NCs also proved to be effective saturable absorbers and realized mode-locking pulses generation found at 1, 1.5 and 2 µm wavebands.[19]

In this paper, $Cu_{3-x}P$ NCs were synthesized using a "one-pot" approach and then drop-casted onto D-shaped fibers as evanescent field interactional saturable absorbers. Through tuning the concentration of $Cu_{3-x}P$ NCs solution, stable mode-locking and Q-switching pulses at 1560 nm can both be generated. The 3 dB bandwidth of the mode-locking optical spectrum is as high as 7.3 nm and the corresponding pulse duration is shortly at 423 fs. The key indicator is that the all fabricated SA devices (over 20 devices) can generate stable mode-locking pulses with similar characteristics, indicating quantum particles material' good repeatability. We therefore believe that heavily-doped colloidal plasmonic NCs are promising and effective SAs for ultra-short pulse generation and other nonlinear photonic applications.

**Material preparation and characterizations**

These self-doped colloidal copper phosphide ($Cu_{3-x}P$) NCs was synthesized by low temperture solution method. All synthesize processes were carried out under standard air-free Schlenk line techniques under nitrogen atmosphere.[14, 20] A solution of 890 mg (2.4mmol) TOP, 60 mg (0.6 mmol) CuCl and 5 ml of OLA was degassed at 100 °C in a 25 mL three-neck flask and kept under vacuum for 1 h. The solution was then heated to a desired temperature under nitrogen (60-180 °C) and 500 mg (0.1mmol) tris(trimethylsilyl)phosphiphine $(TMSi)_3P$ (10 wt% in hexane) was added under vacuum then dissolved in 1 ml ODE to remove the hexane. The $(TMSi)_3P$ solution was then quickly injected into the hot solution. After 3 minutes, the reaction was stopped by removing the heating mantle and purified by precipitating hexane/acetone two times and stored as solid form in a nitrogen-filled glovebox. For the in-situ thermal treatment process, the above reaction need to be heated to 300 °C for 10 min and then the heating mantle was taken out. The purification process is the same as above. The concentration of $Cu_{3-x}P$ NCs that we chose for the experiement is around 5 mg/L.

The as-prepared $Cu_{3-x}P$ NCs was spined coating onto the $SiO_2$ substrate for characterization and the scanning electron microscopy (SEM) was shown in Figure 1. These NCs with hexagonal morphology are densely arranged on the substrate and the size of these NCs is similar. From a single NC as a representative in the inset, the size is around 20.1 nm. In the atomic force microscopy (AFM) image, The thickness of the single nanoplate is around 2 nm. For observing the LSPR effect of the $Cu_{3-x}P$ NCs, the optical absorption spectra were characterized as shown in Figure 1c, where the LSPR peak of the $Cu_{3-x}P$ NCs is around 1456 nm. When spining coating the solution onto the quartz substrate, the LSPR peak occurs a significant redshift to 1562 nm, which is mainly caused by the Coulomb gravitational between nearest-neighbor NCs when forming the film[21]. In order to further ensuring the optical absorption peak in Figure 1c is induced by the LSPR effect other than the bandgap effect, the PL characterization is lauched as shown in Figure d-f. Figure 1d is the optical image of a $Cu_{3-x}P$ NCs assembly and Figure 1e is the corresponding PL mapping results. It is clear that the $Cu_{3-x}P$ NCs have strong PL effect and the PL peak is around 600 nm, which is arised by the optical interband trasition. It is worthy to be noted that the PL peak is relatively broad, which mainly because that the optical bandgap is very sensitive to the NCs size.

**$Cu_{3-x}P$ NCs as SAs for ultrafast fiber lasers**

The laser cavity is the typical ring cavity structure and the working wavelength is around 1550 nm. The components include a Erbium-doped fiber with the length of 0.75 m (LIKKI Er-80/125), a 980 nm/1550 nm wavelength division multiplexer (WDM) pumped by a 976 nm laser diode, a 10% coupler as an output and a polarizer controller (PC) for adjusting the intracavity polarization.

When the brand new side-polished fiber was inserted into the laser cavity without $Cu_{3-x}P$ NCs onside, only continuous wave could be observed unless adjusting the polarization state and the pump power, as shown in Figure S1a. The corresponding optical image of the side polished fiber is posted in the inset with scale bar ~ 50 μm. Then we dropped cast the $Cu_{3-x}P$ NCs solution onto the side-polished fiber as the SA. In this way, $Cu_{3-x}P$ NCs would interact with the intracavity light through the evanescent field effect. It is interesting that the laser output state evolving from continuous wave to mode-locking state with the increase of the amount of $Cu_{3-x}P$ NCs from 20 μL to 120 μL, as shown in Figure S1b - f. During this process, the pump power maintained at 150 mW and no other intracavity parameters would be changed. The evolution of the side-polished fiber coupled SA are also recorded in the insets. The smooth optical spectrum curve in Figure S1f indicates that the stable mode-locking state in which the continuous wave component is suppressed. The clear kelly side-band exhibits the laser was working on the anomalous dispersion state and the 3 dB bandwidth is 3.5 nm.

Then the cavity parameters were carefully tuned, including the cavity length, the intra-cavity dispersion and the polarization states. Typical soliton mode-locking results with high-quality would be achieved as shown in Figure 2. From the optical spectrum in Figure 2a, the center work wavelength is 1570 nm with the 3dB bandwidth broadly at 7.3 nm. The pulse train with the repetition of 31 MHz without clear modulation on the top of the train, indicates the high quality of the mode-locking output (Figure 2b). The single pulse envelop was fitted by $Sech^2$ formula and the pulse duration is shortly at 423 fs, as shown in Figure 2c. The signal-noise rate of the laser is highly at 75 dB, showing the high stability (Figure 2d). Furthermore, from the RF spectrum with a wide frequency range (highly at 1.5 GHz) in the inset of Figure 5d, the output laser pulses are operated stably and no clear noise signal can be observed. By continually running the laser over 6 hours, the optical spectrum still remains stable, further prove that the stability is high (Figure 2e).

When dropping cast over 240 μL $Cu_{3-x}P$ NCs onto the side-polished fiber, the modulation depth, as well as the extra saturable loss of the $Cu_{3-x}P$ NCs based SA would be increased and the typical Q-switching pulses output would be generated. Figure 3a-d shows the Q-switched pulses characteristics under the pump power of 300 mW. In Figure 3a, the central wavelength is 1560.9 nm and the 3-dB spectral bandwidth is more than 1.2 nm, the smooth spectrum envelop indicates that the spectral component of the continuous wave is fully suppressed, illustrating the typical Q-switching state. The Q-switching pulse train in Figure 3b has a relatively high repetition rate of 51.1 kHz and the corresponding time interval between adjacent pulses is 19.5 μs. There is no obvious amplitude modulation (at the fundamental or harmonic frequency of the cavity) on each individual Q-switched pulse envelope and the shape of each pulse have symmetric intensity profile. Single pulse duration is shortly at 1.78 μs (Figure 3c). To study laser stability, a radio frequency (RF) spectrum was also investigated for studying the operation stability, as shown in Figure 3d. The signal-to-noise ratio exceeds 47 dB, which is a quite high value compared to other Q-switched fiber lasers reported in other papers. Furthermore, from the RF spectrum with a wide span range (Inset in Figure 3d), in addition to the fundamental and harmonic frequencies, no any other frequency components can be observed, which confirms that the resulting Q-switched pulses are highly stable. As known that with the increase of the pump power, different output parameters of the Q-switched laser would be changed, including repetition rate, pulse duration, output power and single pulse energy. Figure 3e-f shows the change of these four output parameters with the increase of the pump power at over 560 mW. It is interesting that the repetition rate linearly increases at 88 kHz while the pulse duration decreases dramatically at first and then stabilizes at around 1.3 μs (Figure 3e). Meanwhile, the output power also gradually increases up to 10 mW and the huge Q-switching pulse in principle has large pulse energy. The largest pulse energy is highly at 120 μJ under the pump power of 486 mW. It is noticed that the output quality of the laser remains very stable and no obvious disturbance can be observed under any pump power (provided in Figure S2).

Apart from the excellent performance of the $Cu_{3-x}P$ NCs based mode-locked and Q-switched lasers, the biggest advantage of the solution SAs would probably be the good repeatability and suitability for large scale industrial production. Here we fabricated 10 pieces of $Cu_{3-x}P$ based Q-switched SAs (240 μL $Cu_{3-x}P$ NCs on the side-polished fiber) and 11 pieces of $Cu_{3-x}P$ based mode-locked SAs (120 μL $Cu_{3-x}P$ NCs on the side-polished fiber). All of them effectively generated Q-switching or mode-locking pulses as expected. The optical spectra generated these SAs under the pump power of 300 mW are recorded and both the values of 3dB band width and center wavelength are summarized in Figure 4a and b. The $Cu_{3-x}P$ NCs based mode-locked fiber laser normally have the 3dB band width around 7 nm and the center wavelength around 1569 nm, while these two values in Q-switched fiber lasers are around 1.2 nm and 1561 nm, respectively. The slight difference of the output parameters in different SAs may be caused by the inaccurate fabricating process, for example, the fiber core is so small that many $Cu_{3-x}P$ NCs are aligned on the Fiber cladding and do not join in the interaction of the intra-cavity light. Overall, the investigation of over 20 $Cu_{3-x}P$ based SAs provides the direct evidence that the new kind of SA has promising potential for large-scale industrial production.

**Conclusion**

In this paper, $Cu_{3-x}P$ NCs as promising SA materials are employed for ultrafast pulse generation. Through tuning the amount of $Cu_{3-x}P$ NCs solution, both mode-locking and Q-switching pulses with high quality at around 1550 nm can be generated. The 3 dB bandwidth of the mode-locking optical spectrum is observed as high at 7.3 nm and the corresponding pulse duration is shortly at 423 fs. The repetition rate of the Q-switching pulses is high than 80 kHz and the largest pulse energy is more than 120 µJ. Most noticeably, all fabricated SA devices (over 20 devices) can generate stable mode-locking pulses as well as Q-switching pulses with similar characteristics, indicating that the quantum particles material has good reproduction for fabricating SA devices and promising potential for large-scale industrial production. We believe that heavily-doped colloidal plasmonic NCs are promising and effective SAs for ultra-short pulse generation and other nonlinear photonic applications.

**Acknowledgments**

We acknowledge the support from the National Natural Science Foundation of China (No. 61875139, 91433107 and 51702219), the National Key Research & Development Program (No. 2016YFA0201902), Shenzhen Nanshan District Pilotage Team Program (LHTD20170006) and Australian Research Council (ARC, FT150100450, IH150100006 and CE170100039). B.S. acknowledges the funding support from China Postdoctoral Science Foundation Grant (No. 217M622758).

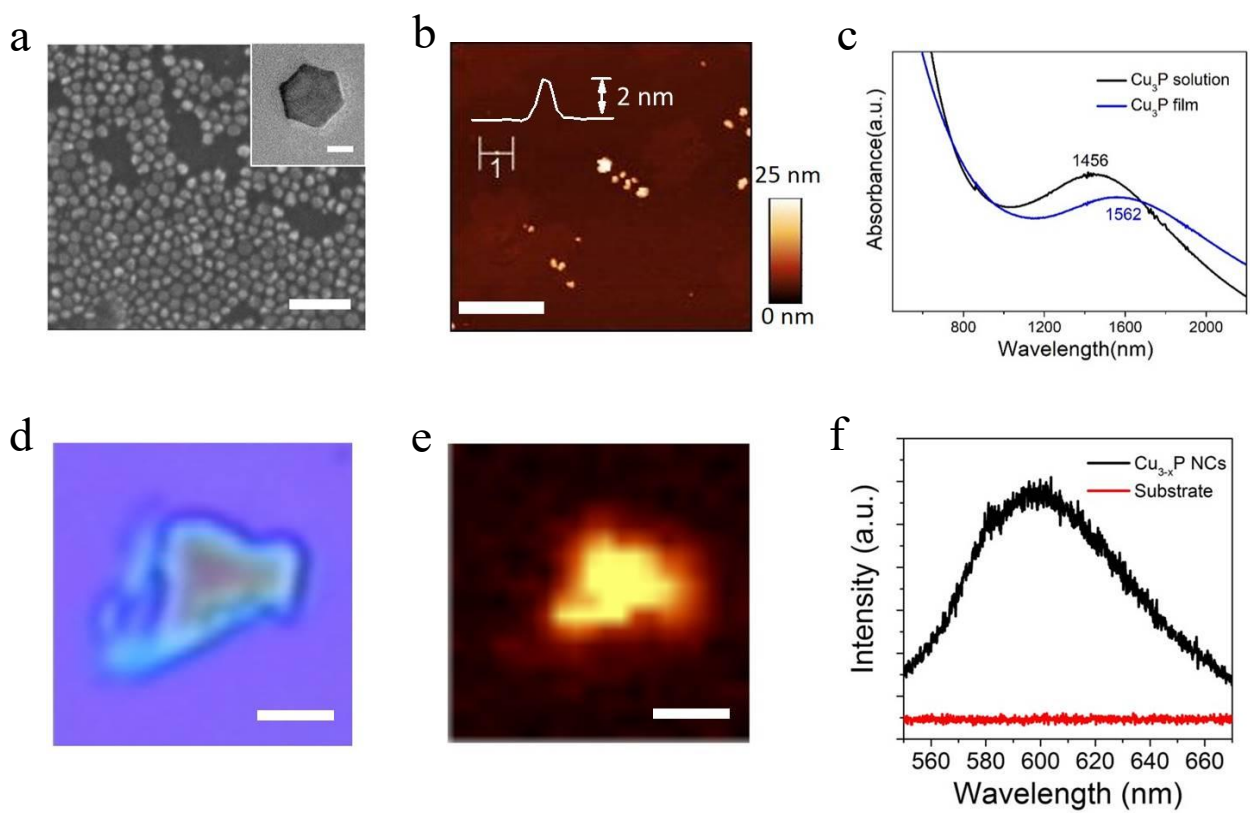

**Figure 1**. Material morphology and optical characterizations of $Cu_{3-x}P$ NCs. a) SEM image of $Cu_{3-x}P$ NCs (Scale bar: 100 nm) and TEM image of the single $Cu_{3-x}P$ NC (inset, scale bar: 10 nm). b) AFM image (Scale bar: 1 µm). c) Optical absorption spectra of $Cu_{3-x}P$ NCs in toluene solution (Black line) and on silicon substrate (Blue line), respectively. d) Optical image of $Cu_{3-x}P$ NCs assembly and e) the corresponding PL mapping results (Scale bar: 1 µm). f) PL spectrum of $Cu_{3-x}P$ NCs assembly.

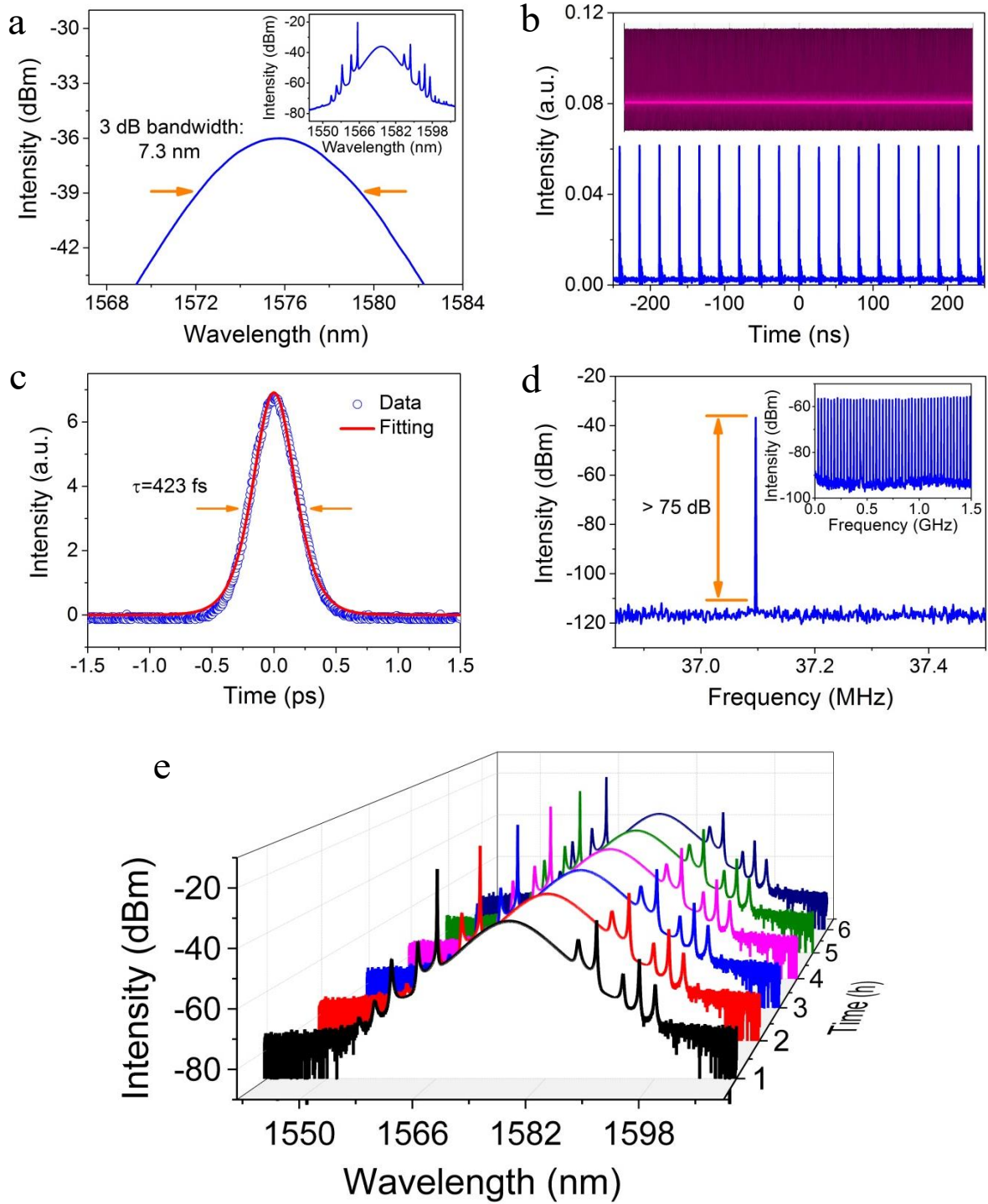

**Figure 2.** Typical mode-locking characteristics. a) Typical mode-locking optical spectrum. b) Mode-locking pulse train. c) Autocorrelation trace. d) The RF optical spectrum at the fundamental frequency and the wideband RF spectrum (inset). e) Long term stability of the optical spectra measured under different time.

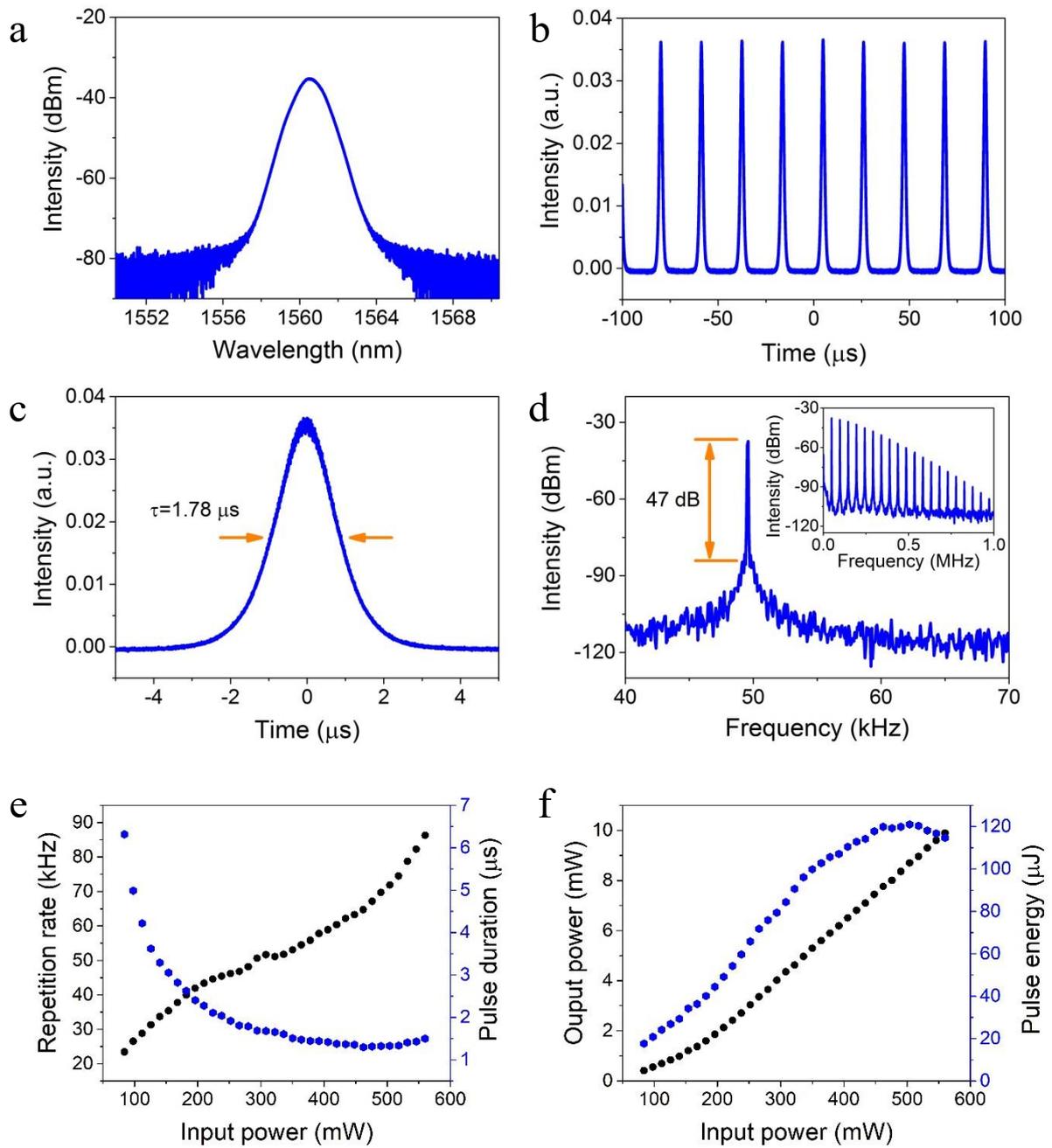

**Figure 3.** Q-switching pulse output characterizations. a) Optical spectrum. b) Q-switching pulse train. c) Single Q-switching pulse. d) The radiofrequency optical spectrum at the fundamental frequency and the wideband RF spectrum (inset). e) Pulse repetition rate and duration versus incident pump power. f) Output power versus incident pump power.

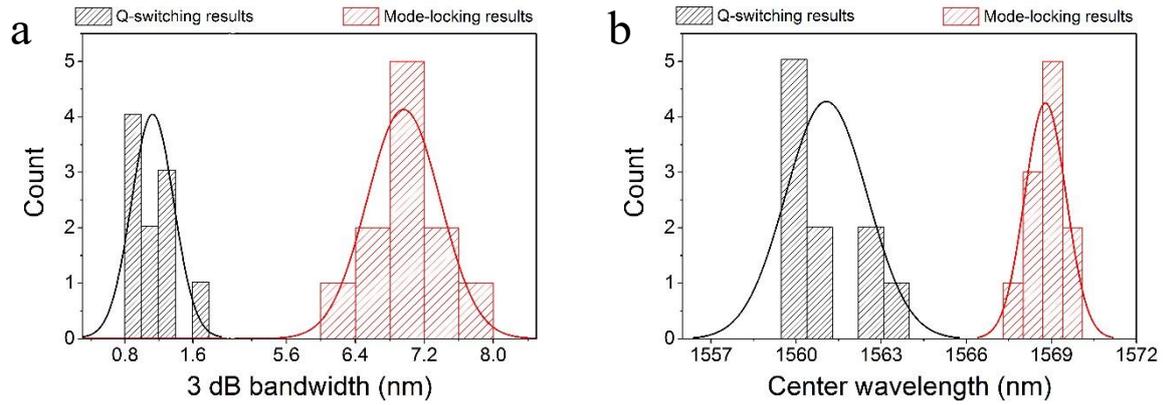

**Figure 4.** Statistical distribution of optical spectra characteristics of 10 pieces of $Cu_{3-x}P$ based Q-switched fiber lasers (Black marked) and 11 pieces of $Cu_{3-x}P$ based mode-locked fiber lasers (Red marked). (a) Statistical distribution of 3 dB bandwidth and corresponding Gauss-fitting results. (b) Statistical distribution of center wavelength and corresponding Gauss-fitting results.

# Supporting information

**Evanescent Field Functional $Cu_{3-x}P$ Nanoparticles as Effective Saturable Absorbers with high Repeatability for Femtosecond Soliton Pulse generation**

*Haoran Mu[1], Zeke Liu[2], Zhichen Wan[1], Barbar Shabbir[1], Lei Li[3], Tian Sun[2], Shaojuan Li[2], Wanli Ma[2], and Qiaoliang Bao[1,*].*

[1]Department of Materials Science and Engineering, Monash University, Clayton, Victoria 3800, Australia.

[2]Institute of Functional Nano and Soft Materials (FUNSOM), Jiangsu Key Laboratory for Carbon-Based Functional Materials and Devices, and Collaborative Innovation Center of Suzhou Nano Science and Technology, Soochow University, Suzhou 215123, P. R. China.

[3]Jiangsu Key Laboratory of Advanced Laser Materials and Devices, Jiangsu Collaborative Innovation Center of Advanced Laser Technology and Emerging Industry, School of Physics and Electronic Engineering, Jiangsu Normal University, Xuzhou, Jiangsu 221116, China

*Address correspondence to (Q. Bao) Qiaoliang.Bao@ monash.edu

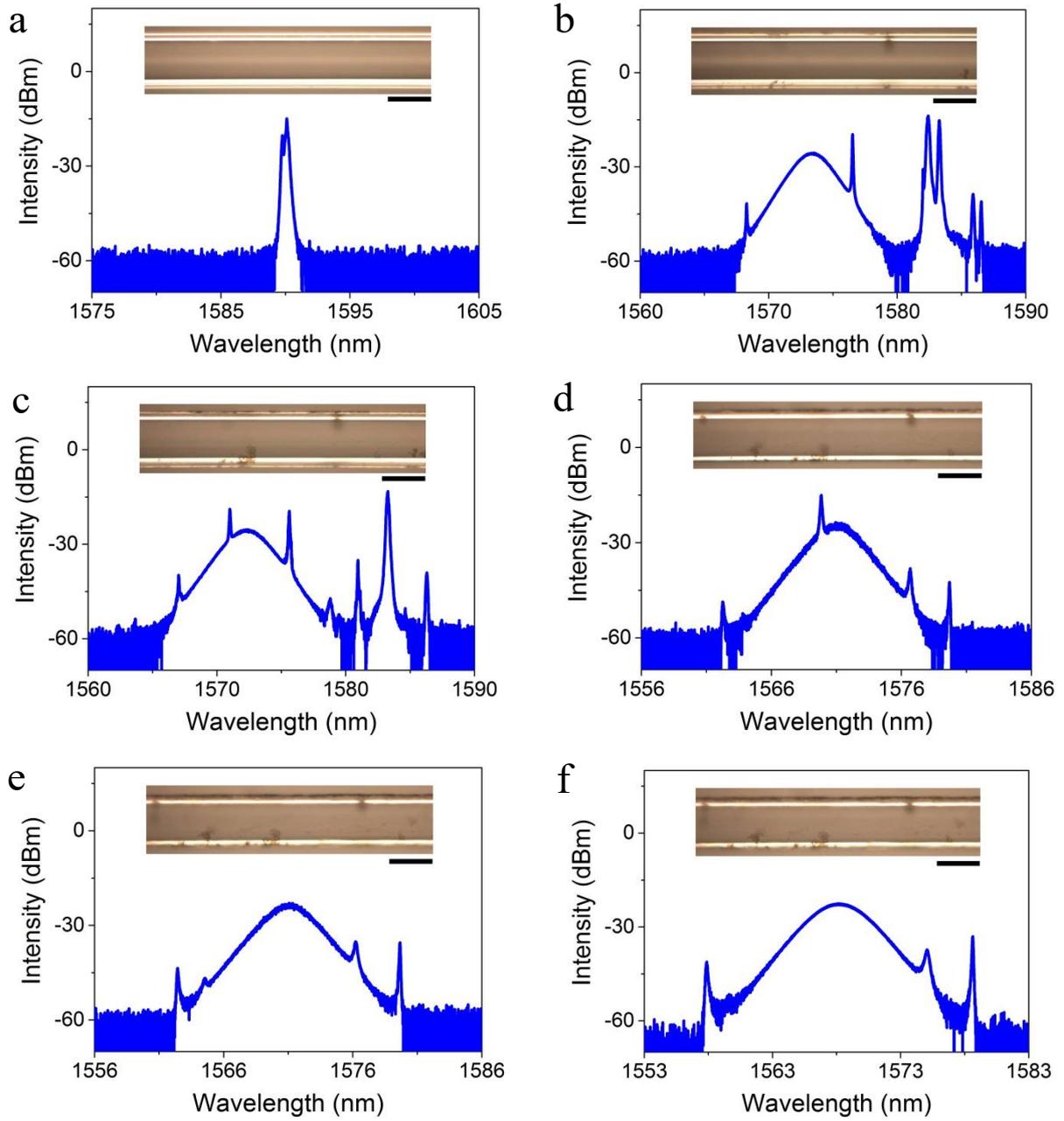

**Figure S1.** evolution of output optical spectra with the increase of $Cu_{3-x}P$ solution under the same condition of intra-cavity polarization state and pump power (scale bar: 100 μm). a) 20 μL. b) 40 μL. c) 60 μL. d) 80 μL. e) 100 μL. f) 120 μL.

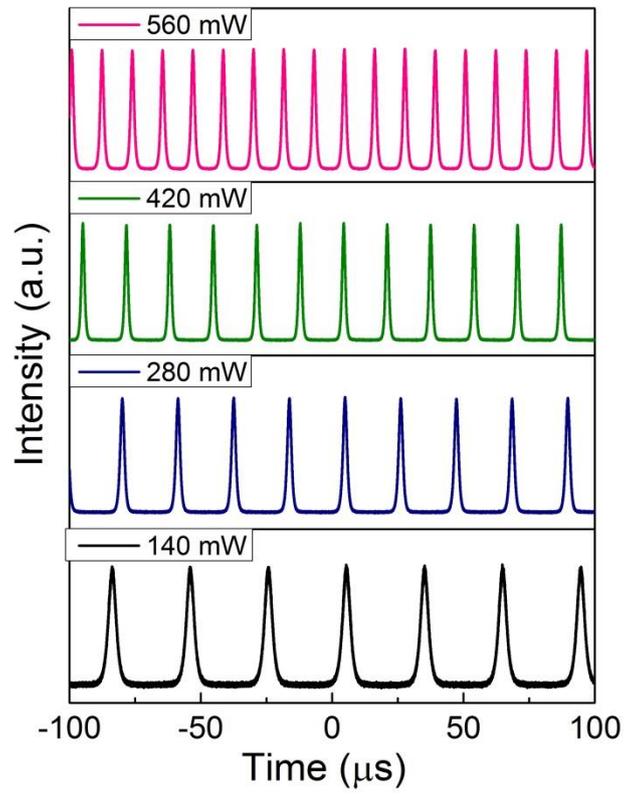

**Figure S2.** Various Q-switching pulse trains obtained at different pump powers.